\documentclass[12pt]{iopart}

\usepackage{graphicx}
\usepackage[usenames]{color}

\begin{document}

\title[Atom Phase Lock]{Locking Local Oscillator Phase to Atomic Phase via Weak Measurement}

\author{N Shiga$^{1, 2}$ and M Takeuchi$^3$}

\address{$^1$ PRESTO, Japan Science and Technology (JST), 4-1-8 Honcho Kawaguchi, Saitama 332-0012, Japan}
\address{$^2$ National Institute of Information Communication and Technology, 4-2-1 Nukui-kitamachi, Koganei, Tokyo 184-8795, Japan}
\address{$^3$ Department of Basic Sciences, Graduate School of Arts and Sciences, the University of Tokyo, 3-8-1 Komaba, Meguro-ku, Tokyo 153-8902 Japan}
\ead{shiga@nict.go.jp}

\begin{center}
\today
\end{center}

\begin{abstract}
A new method is proposed to reduce the frequency noise of a local oscillator to the level of white phase noise by maintaining (not destroying by projective measurement) the coherence of the ensemble pseudo-spin of atoms over many measurement cycles.
This method, which we call ``atomic phase lock (APL),'' uses weak measurement to monitor the phase in the Ramsey method and repeat the cycle without initialization of the phase.  APL will achieve white phase noise as long as the noise accumulated during dead time and the decoherence are smaller than the measurement noise.
A numerical simulation confirmed that with APL, Allan deviation is averaged down at a maximum rate that is proportional to the inverse of the total measurement time, $\tau^{-1}$.
In contrast, current atomic clocks that use projection measurement suppress the noise only to the white frequency noise level, in which case the Allan deviation scales as $\tau^{-1/2}$.
Faraday rotation is one way to achieve weak measurement for APL.  The strength of Faraday rotation with $^{171}\rm{Yb}^+$ ions trapped in a linear rf-trap is evaluated, and the performance of APL is discussed.
The main source of the decoherence is a spontaneous emission, induced by the probe beam for Faraday rotation measurement.
The Faraday rotation measurement can be repeated until the decoherence becomes comparable to the signal-to-noise ratio of the measurement.  The number of cycles for a realistic experimental parameter is estimated to be $\sim$100.
\end{abstract}


\section{Introduction}
Many applications and experiments use the electromagnetic (EM) field to manipulate the quantum state of two-level systems (TLS).
Improving the frequency stability of the EM field and TLS are of great importance because
the ability to precisely and coherently control the population and phase of TLS plays a crucial role in many applications, such as nuclear magnetic resonance (NMR) spectroscopy and imaging, atomic clocks, magnetometers, quantum computers, and quantum simulators.
The frequency (or phase) of the EM field is usually referenced to the oscillator called the local oscillator (LO).

A no-feedback approach, known as spin echo, has been developed in NMR to suppress the dephasing error \cite{Hahn1950}.
The spin echo uses the phase of the LO as a reference and averages out the phase error of TLS by inserting $\pi$-pulses.
A similar technique can also be applied to isolate non-classical effects, such as entanglement with the noise environment, and is called ``dynamical decoupling'' \cite{Viola1998}.
No-feedback approaches are very useful and easy to implement because the phase difference between LO and TLS does not need to be monitored.  However, it does not suit applications such as atomic clocks and magnetometers, whose goal is to stabilize LO by using TLS as a reference.  Spin echo also cannot maintain the phase error in the long term.
Since we are pursuing the long-term stability of LO, we chose to use the feedback approach.
Different methods are used depending on the kind of target and reference oscillators and whether the noise is suppressed to white frequency noise or to white phase noise.  A summary of various feedback methods is shown in \Tref{tab:VariousMatching}.

\begin{table}[h]
  \centering
\begin{tabular}{|c|c||c|c|}
  \hline
  \textbf{Target}    & \textbf{Reference} & \textbf{White frequency noise} & \textbf{White phase noise}    \\
     &     & $\sigma_y\propto\tau^{-1/2}$  & $\sigma_y\propto\tau^{-1}$  \\ \hline\hline
  Laser & Laser(LO) & Transfer cavity & Laser phase lock \\ \hline
  TLS   & LO    & NMR lock \cite{Vanzijl1987}   & Coherent magnetometry \cite{Stockton2004a} \\ \hline
  LO & TLS & (Conventional) atomic clock   & ``Atomic phase lock'' \\ \hline

\end{tabular}
  \caption{Various ways to match the frequencies of oscillators.  $\tau$ is the total measurement time, and $\sigma_y$ is the Allan deviation (deviation of the target frequency from the reference frequency) of the normalized frequency (explained in Section \ref{ch:Allan}).
When both the target and reference oscillators are lasers, white frequency noise is often achieved by locking the target laser to the reference laser by matching the frequency of the resonant peaks.
For example, the transfer cavity length is locked to the reference laser frequency, and then the target laser frequency is locked to one of the transfer cavity's resonances.
The NMR lock looks at the NMR signal of the deuterated solvents, and the magnetic field is feedback controlled to keep the atomic spin resonant frequency constant \cite{Vanzijl1987}.
An atomic clock is a typical example of an LO frequency being locked to the atomic spin resonance frequency.
In both cases, noise was suppressed to white frequency noise, and $\sigma_y$ was reduced at a rate of $\tau^{-1/2}$ as derived in Section \ref{RamseyPhase}.
}\label{tab:VariousMatching}
\end{table}

Conventional atomic clocks use the Ramsey method with projection measurement.  Although this Ramsey method compares the phases of LO and TLS,  it cannot achieve white phase noise over many measurement cycles because the phase of the atomic spin is destroyed and initialized at each cycle due to projection measurement.
As these measurement cycles are repeated, the measurement noise accumulates at each cycle and the phase uncertainty grows as $\delta_\phi\propto\tau^{1/2}$.
As a result, frequency stability decreases at a rate of $\sigma_y\propto \tau^{-1/2}$, which is characteristic of white frequency noise.
To achieve white phase noise, we need to monitor the phase of the atomic spin over many cycles without destroying it.

We propose an experimental method called ``atomic phase lock (APL)'' to achieve white phase noise.
This method combines weak measurements \cite{Aharonov1988, Yokota2010} with the Ramsey method to monitor the phase difference while having the least effect on the coherence of the spin over many cycles.

With APL, $\sigma_y$ can be reduced at a faster rate up to  $\tau^{-1}$ as noise is suppressed to white phase noise.
This $\sigma_y\propto \tau^{-1}$ is achieved when the phase of the target oscillator is locked to the phase of the reference oscillator.
Although white phase noise is routinely observed when locking a target laser phase to a reference laser phase, feedback control of phase matching between TLS and LO has not been achieved.
This is because TLS is a passive oscillator and monitoring the phase of TLS is difficult without affecting (destroying) the phase itself.

This paper is organized as follows.
In Section 2, atomic clocks that run with projection measurement are reviewed, and APL is introduced.
In Section 3, we estimate the Allan variance for atomic clocks with projection measurement and with APL.
In Section 4, we propose the $^{171}\rm{Yb}^+$ ion trap with microwave transition as a clock transition for demonstration of the proof-of-principle.
Then, the signal strength of Faraday rotation is estimated.  The decoherence rate is calculated to estimate how many cycles of Faraday rotation measurement can be performed.
In Section 5, we compare the stability of different types of atomic clocks using numerical simulation.
In Section 6, we discuss
systematic shifts of the $^{171}\rm{Yb}^+$ microwave clock,
a comparison between APL and the spin-squeezed Ramsey clock, and
the extension of APL to the optical clock transition.

\section{\label{ch:APL}Ramsey method and APL}

\subsection{\label{RamseyPhase}Review of Ramsey method with projection noise}
The phase of the spin (relative to the LO phase) cannot be measured directly. The Ramsey double pulse method, in essence, measures the spin phase by mapping the phase information to the population ratio. Projection measurement of the population ratio gives us the phase.  Note that the Ramsey method is often interpreted as a measurement of the average frequency shift during the time between double pulses.  We instead view the Ramsey method as a measurement of the phase accumulated during the time between two pulses.  This is because viewing it as a phase measurement is important for understanding the advantage of our proposed APL.  If the Ramsey signal is expressed as a function of frequency, $\phi=\int \Delta\omega(t)\ dt$ can be used to change the dependence on $\phi$.

Conventionally, the phase information represented by the population ratio has been measured via projection measurement.  The Ramsey method with projection measurement is called ``projection Ramsey'' in this paper.  This is to clarify the difference to APL, whose signal detection is via weak measurement.

In this section, we will review the projection Ramsey method using Bloch representation \cite{Bloch1946}.
The dynamics of a two-level system interacting with an EM field are the same as the dynamics of a spin-$\frac{1}{2}$ particle in a magnetic field \cite{Feynman1957}, and we will only use this ``spin'' picture in the rest of this paper.
The Bloch representation is useful for discussing the time evolution of the spin interacting with the field.
For a more detailed explanation of the Bloch sphere in the context of atomic clocks, see, for example, Ref.\cite{Riehle2004}.

\begin{figure}[htbp]
\begin{center}
  \includegraphics[width=160mm]{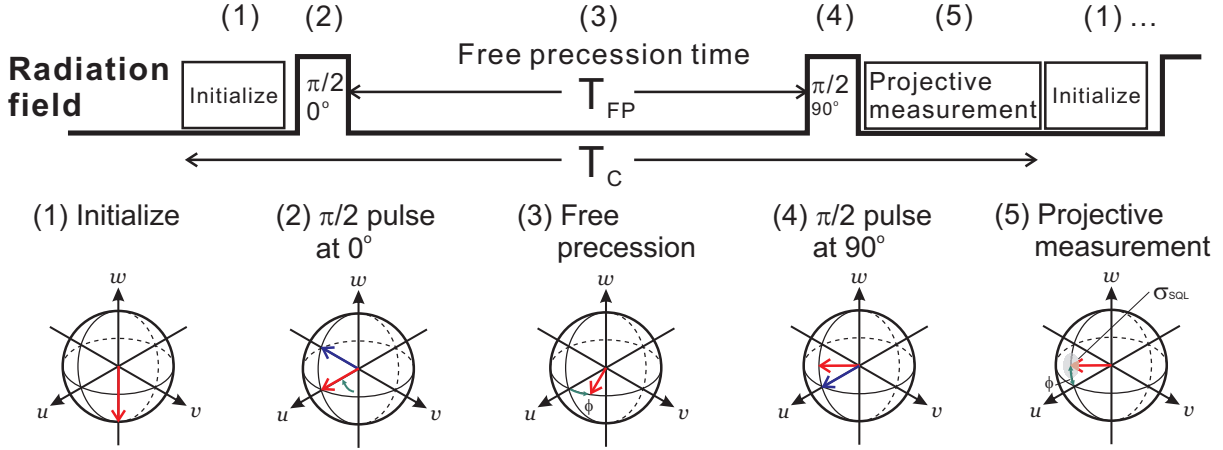}\\
  \caption{Step-by-step description of the Ramsey method with Bloch sphere pictures.  The red arrow is the spin vector, and the blue arrow is the torque vector that rotates the spin. The angle in degrees is the phase of the EM field, and the angle in radians is the rotation angle of the spin by the EM field.  $T_C$ is the cycle time.
}\label{fig:Ramsey}
\end{center}
\end{figure}

The projection Ramsey sequence is shown in \Fref{fig:Ramsey}.
The Cartesian coordinates of the Bloch sphere are designated as the \emph{u, v,} and \emph{w} axes.
We limit our argument to the case where the phase of the spin is coherent (pure state).
The ground and excited state correspond to the south and north poles respectively, and the phase of the spin corresponds to longitude.  We use the frame that rotates at $\omega_{LO}$ and the rotating-wave approximation.  Then we consider the following two types of rotation.
A spin is rotated around a torque vector that lies on the $u-v$ plane when the resonant EM field is applied.
A spin is also rotated along the longitudinal direction at a rate given by $\Delta \omega$, which is the frequency difference of LO and atoms.
For clarity, we will express the phase of the EM field in degrees and the rotation angle in radians.

\begin{figure}
  \begin{center}
   \includegraphics[width=70mm]{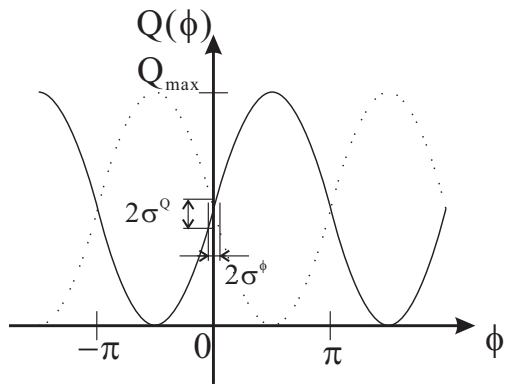}\\
   \caption{Graph of the mean detected signal as a function of the phase difference accumulated during $T_{FP}$ in the Ramsey method.  The solid and dotted lines show cases of $+90^\circ$ and $-90^\circ$ phase shifts respectively in step (4) in \Fref{fig:Ramsey}.
   }\label{fig:RamseySpectrum}
  \end{center}
\end{figure}

Ramsey measurement proceeds as follows.
(1) Repump all the atoms to $|g\rangle$ to correspond to aligning the spin to the coherent spin state pointing along $-\emph{w}$.  (2) Apply the strong resonant EM field to rotate the spin around $\emph{-v}$ by $\pi/2$ to point $\emph{u}$.  We define this phase of the EM field (LO) as $0^\circ$ and assume the power is strong enough that the time taken for this rotation is negligible.  (3) Wait for free precession time $T_{FP}$, and the frequency difference between LO and the atom rotates the spin around $\emph{w}$.  The angle $\phi$ between $\emph{u}$ and the spin corresponds to the phase difference of LO and atomic spin that is accumulated during $T_{FP}$.  (4) Apply the second EM field with $90^\circ$ phase shift to rotate the spin around $\emph{u}$ by $\pi/2$.  Now the angle $\phi$ is represented by the value along $\emph{w}$, which is a population ratio of the superposition state.  Again, we assume that a negligible amount of time is taken for this step.  (5) Projection measurement signal $Q$ gives the measure of the population ratio between $|g\rangle$ and $|e\rangle$.
The mean detected signal of a $Q$ is shown in \Fref{fig:RamseySpectrum}.  Signal $Q(\phi)$ is a sinusoidal function of the phase $\phi$ and is given by
\begin{equation}\label{eq:RamseySpectrum}
    Q(\phi)=\frac{Q_{max}}{2}(1+\sin(\phi))
\end{equation}
The passive atomic clock uses this slope near $\phi=0$ in \Fref{fig:RamseySpectrum} as an error signal to lock the frequency.

The projection measurement in step (5) is normally done with electron shelving, first proposed by Dehmelt \cite{Dehmelt1975, Nagourney1986}.  Applying a laser beam appropriately polarized and tuned to a transition will scatter many photons if the atom is in one of the two level's states (a ``cycling" transition) but will scatter no photons if the atom is in the other state.  The population ratio can be measured by counting the number of bright ions.
The value of the population ratio has a fundamental fluctuation of the quantum projection noise (QPN) \cite{Itano1993}.  QPN originates from the randomness in quantum projection and follows the binomial distribution (for the mathematical expression, see Eq. \eref{eq:QPN}).   This QPN is then reflected in the measurement of $\phi$ through Eq. \eref{eq:RamseySpectrum}.

We shift the phase of the EM field by $90^\circ$ to obtain the greatest sensitivity (largest slope) at around $\phi=0$.  Applying two EM field pulses with the same phase is common, and this minor modification of the phase itself will not affect the performance of the atomic clock.

The atomic clock's performance is normally improved by extending the free precession time ($T_{FP}$) of the Ramsey method, but $T_{FP}$ is limited by the stability of the LO.
In other words, the $T_{FP}$ can be longer only as long as the phase difference accumulated in $T_{FP}$ is guaranteed to be within $\pm \pi/2$.
In the aim for a longer $T_{FP}$ time, the stability of the laser (LO) has been improved by locking the laser frequency to a high-finesse cavity with ultra-low-expansion (ULE) glass as a spacer.  However, the thermal noise of the mirror coating \cite{Numata2004} is a hard limit to break through with the present technology, and improving the $T_{FP}$ by an order of magnitude would be difficult.
In such a situation, APL could provide the alternate path to break through this limit to achieve a longer $T_{FP}$ because APL is equivalent to lengthening the $T_{FP}$, as long as the atom coherence is maintained.

\subsection{\label{AtomPhaseLock}Atomic phase lock}
\begin{figure}[htbp]
\begin{center}
  \includegraphics[width=160mm]{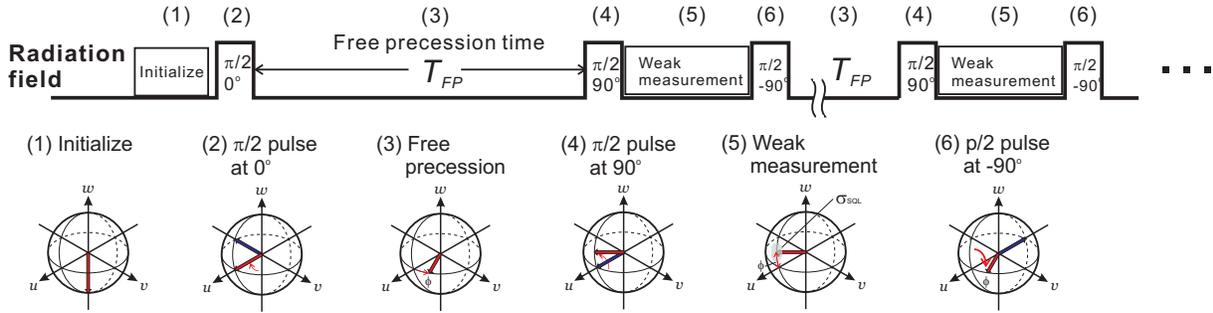}\\
  \caption{Step-by-step description of APL with Bloch sphere pictures.
}\label{fig:AtomPhaseLock}
\end{center}
\end{figure}

The APL method we propose is structured as follows (\Fref{fig:AtomPhaseLock}). (1) Initialize the spins along $-\emph{w}$. (2) Rotate the spin by a $\pi /2$ pulse with 0$^\circ$ phase. (3) Wait for the free precession time $T_{FP}$.  (4) Rotate the spin by a $\pi /2$ pulse with 90$^\circ$ phase. (5) Estimate the +$\emph{w}$ component via the weak measurement. (6) Rotate back the spin by a $\pi /2$ pulse with -90$^\circ$ phase. (7) Repeat steps (3) to (6).

The differences from the projection Ramsey method are that the projection measurement is replaced by weak measurement and that, after the weak measurement, the spin is rotated back to where it was before step (4) and the process returns to step (3) without initialization.
The point of APL is to monitor the phase without destroying the coherence of the spin so that the measurement noise does not accumulate after many cycles.
A dispersion measurement serves as a weak measurement, and there are two ways to perform it: Mach-Zender interferometry \cite{Lodewyck2009} and Faraday rotation \cite{Takahashi1999}.  In Section \ref{FaradayRotation}, we will discuss the Faraday rotation on spins of trapped ions.

\section{\label{ch:Allan}Allan deviation and stability of atomic clocks}
In this section, we will estimate the frequency stability of atomic clocks.  For this, we use the Allan deviation, a widely used measure for evaluating the frequency noise of the LO, which is in general non-stationary and correlated.
We first define terms and symbols.  Fractional frequency deviation is defined as
\begin{equation}\label{}
    y(t)\equiv\frac{\Delta\nu(t)}{\nu_0}=\frac{\nu(t)-\nu_0}{\nu_0},
\end{equation}
where $\nu_0$ is the clock frequency of atoms and assumed to be a constant.  The Allan deviation of the fractional frequency $\sigma_y(\tau)$ is defined as
\begin{equation}\label{}
    \sigma_y(\tau)\equiv\sqrt{\frac{1}{n}\sum_{i=1}^n(\overline{y_{i+1}}-\overline{y_i})^2}.
\end{equation}
Here, $\overline{y}$ means the time average of $y$ over time $\tau$.  The Allan deviation is the deviation from the previous data rather than from the mean.


Now, we will estimate the Allan deviation of the atomic clock with the Ramsey method, using \Fref{fig:RamseySpectrum}.  We follow the basic argument in Ref. \cite{Riehle2004} with a modification to view the Ramsey method as phase measurement.
When there is a measurement uncertainty $\sigma^Q$, the corresponding uncertainty in $\phi$, $\sigma^\phi$ is linked by the slope of the spectrum as follows:
\begin{equation}\label{eq:sigma1}
    \frac{dQ}{d\phi}=\frac{\sigma^Q}{\sigma^\phi}.
\end{equation}
From Eq. \eref{eq:RamseySpectrum}, the slope at $\phi=0$ is
\begin{equation}\label{eq:slope}
    \frac{dQ}{d\phi}=\frac{Q_{max}}{2},
\end{equation}
and Eq. \eref{eq:sigma1} is given as
\begin{equation}\label{eq:sigma2}
    \sigma^\phi=\sigma^Q\frac{2}{Q_{max}}=\frac{2}{\rm{SNR}},
\end{equation}
where SNR$\equiv Q_{max}/\sigma^Q$ is the signal-to-noise ratio of a single measurement.

We define $\phi(\tau)$ as the total phase difference between LO and spin after time $\tau$.  In contrast, $\Delta\phi$ represents the phase difference accumulated during the free precession time of the projection Ramsey method.
Allan deviation $\sigma_y(\tau)$ ($\tau$ is total measurement time) is linked to the RMS deviation $\delta_\phi(\tau)\equiv\sqrt{\langle[\phi(\tau)-\langle\phi(\tau)\rangle]^2\rangle}$ as follows:
\begin{equation}\label{eq:Allan1}
    \sigma_y(\tau)=\frac{\delta_\phi(\tau)}{2 \pi \nu_0 \tau}.
\end{equation}

The $\tau$ dependence of $\delta_\phi(\tau)$ is characterized by the type of noise (i.e., white frequency noise, white phase noise, etc.).
For example, the repetition of the projection Ramsey measurement and feedback results in random walk phase noise because the spin phase is randomized by decoherence due to the projection measurement.
Random walk phase noise is also called white frequency noise, and its deviation grows as $\delta_\phi(\tau)\propto\sqrt{\tau}$.  This growth of the noise is due to ``resetting (projection and initialization)'' of the spin phase at each Ramsey cycle.  In this case, $\phi(\tau)$ can be estimated as a sum of the phase measurement at cycles,
\begin{equation}\label{}
    \phi(\tau)=\sum_{i=0}^{N_c}\Delta\phi_i,
\end{equation}
where $N_c$ is the number of cycles given by $N_c\sim\tau/T_c$ and $T_c$ is the time taken for a single Ramsey measurement cycle.
Since $\Delta\phi_i$ are non-correlated due to resetting of the phase at each cycle, the ensemble average of the deviation $\delta_\phi(\tau)$ is expressed using Eq. \eref{eq:sigma2} as
\begin{equation}\label{}
    \delta_\phi(\tau)=\sqrt{\langle[\phi(\tau)-\langle\phi(\tau)\rangle]^2\rangle}
    =\sqrt{\langle[\phi(\tau)]^2\rangle}
    =\sigma^\phi\sqrt{N_c}\sim\frac{2}{\rm{SNR}}\sqrt{\frac{\tau}{T_c}}.
\end{equation}
Here we consider only the case where $\tau>T_C$, and we used $\langle\phi(\tau)\rangle=0$ because the LO frequency does not drift when it is stabilized to an atom.
This growth of the deviation is analogous to the case of measuring a 400 m race track with a 10 cm scale stick.  Every time 10 cm is measured and added to the total length, noise is added due to the finite width of the marker line.
Finally, we get the Allan deviation of the atomic clock with the Ramsey cycle as
 \begin{eqnarray}\label{eq:AllanWhiteFreq}
   \sigma_y(\tau) &=& \frac{1}{\pi \nu_0 \tau \rm{SNR}}\sqrt{\frac{\tau}{T_c}} \\
                  &=& \frac{1}{\pi \rm{SNR}}\frac{\Delta\nu}{\nu_0}\sqrt{\frac{T_c}{\tau}},
 \end{eqnarray}
where $\Delta\nu\equiv1/T_{FP}\sim1/T_c$.  Equation \eref{eq:AllanWhiteFreq} is the same as in Ref. \cite{Riehle2004}, and we call this the ``white frequency noise limit.''  When the SNR is limited by the quantum projection noise SNR$_{QPN}=1/\sqrt{N_a}$ ($N_a$ is the number of atoms), Eq. \eref{eq:AllanWhiteFreq} is sometimes called the ``QPN limit.''


From Eq. \eref{eq:AllanWhiteFreq}, we can see that $\sigma_y(\tau)$ decreases as $1/\sqrt{T_C}$.  Thus, it is better to use a longer free precession time, but normally, the length of $T_{FP}$ is limited by the noise of LO.  In other words, when $T_{FP}$ is too long, the phase difference after $T_{FP}$ becomes larger than $\pm\frac{\pi}{2}$, and whether the signal is, for example, $1/4\pi$ or $3/4\pi$ cannot be distinguished (see \Fref{fig:RamseySpectrum}).  Therefore, in the regular Ramsey sequence, $T_{FP}$ is kept sufficiently small that the phase signal is guaranteed to be within $\pm\frac{\pi}{2}$.

APL aims to overcome this limitation due to the noise of LO by use of the weak measurement.
In essence, APL "peeks" at the phase with the least amount of decoherence, and feedback will keep the phase within $\pm\frac{\pi}{2}$.  As long as the coherence of the spin is sufficiently maintained, the APL achieves a single Ramsey measurement with long $T_c$, that is equal to the total measurement time $\tau$.
As a result, $T_c=\tau$ is achieved and Eq. \eref{eq:AllanWhiteFreq} now becomes,
\begin{equation}\label{eq:AllanWhitePhase}
    \sigma_y(\tau)=\frac{1}{\pi \nu_0 \tau \rm{SNR}}
    \propto\tau^{-1}.
\end{equation}
We call this ideal case of the APL as the ``white phase noise limit'' in this paper.

\section{\label{FaradayRotation}Possible experimental setup to achieve atomic phase lock}

\subsection{Experimental setup}
We will first specify an experimental setup for a quantitative discussion.  We chose a microwave atomic clock, using $^{171}\rm{Yb}^+$ ions trapped in a linear rf-trap, as a possible proof-of-principle experiment.
We will see the white phase noise limit until decoherence degrades the performance of APL.  We chose an ion trap for its long trapping time and chose the magnetic transition of the ground-state hyperfine splitting as a clock transition for its long life time and long coherence time.
For weak measurement, we will focus on Faraday rotation because it is somewhat easier to set up in an experiment.
For simpler design and implementation of the experiment, we chose Ytterbium 171 ions that had the smallest nuclear spin of $1/2$.

\begin{figure}[h]
  \begin{center}
    \includegraphics[width=120mm]{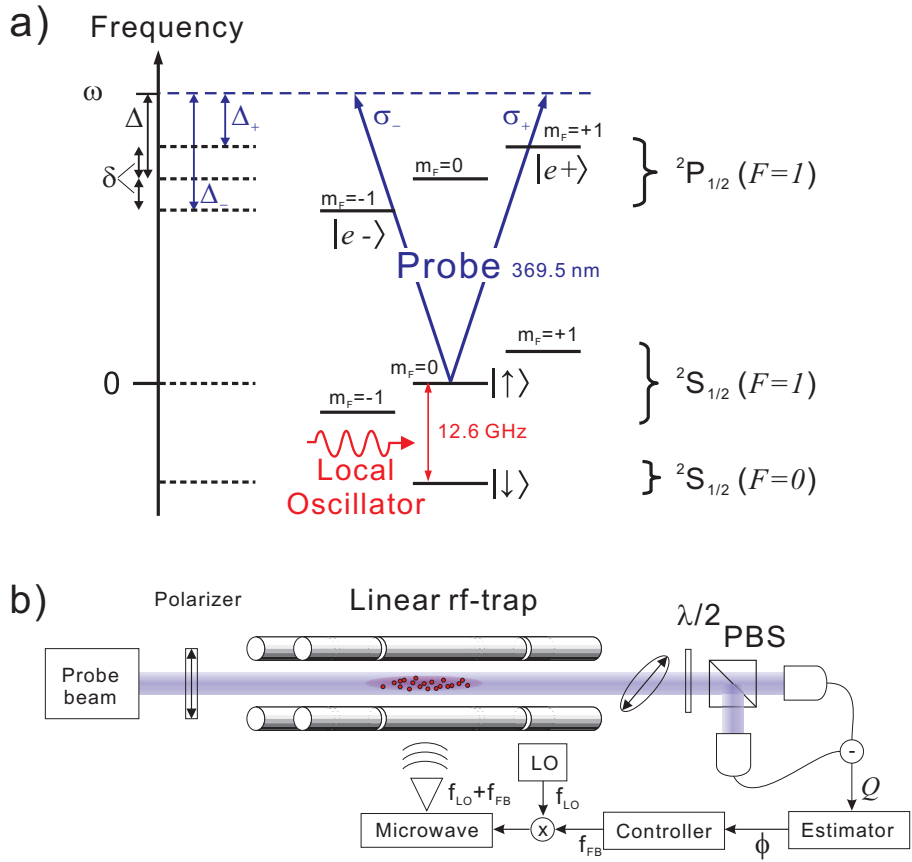}
  \end{center}
  \caption{(a) Energy diagram of $^{171}\rm{Yb}^+$, for the Faraday rotation measurement.  A probe beam for the cooling transition is shown. $|e\pm\rangle$ denotes the excited-state sublevels of dipole-allowed optical transition. (b) Experimental setup for Faraday rotation measurement and the schematics of the feedback.}
  \label{fig:ExperimentalSetup}
\end{figure}

An energy diagram of the Faraday rotation experiment setup is shown in \Fref{fig:ExperimentalSetup}(a).  We assigned $^2S_{1/2}(F=0, m_f=0)$ as the spin-down state $|\!\!\!\downarrow\rangle$ and $^2S_{1/2}(F=0, m_f=0)$ as the spin-up state $|\!\uparrow\rangle$.  We use the transition between these two states as the clock transition.

We assume that one of the clock levels $|\!\uparrow\rangle$ has two dipole-allowed optical transitions of
 $|\!\uparrow\rangle\leftrightarrow|e\pm\rangle$.
This structure is common for the species of atomic clock including the optical frequency domain.
\Fref{fig:ExperimentalSetup}(b) shows a setup for Faraday rotation measurement.
The probe beam is linearly polarized and off-resonant with the detuning of $\Delta_\pm$ for $|\!\uparrow\rangle\leftrightarrow|e\pm\rangle$ transition, respectively.
The probe beam uses the $^2S_{1/2}\leftrightarrow^2P_{1/2}$ transition, whose wavelength $\lambda$ is $\lambda=369.5~$nm, resonant atomic scattering cross-section $\sigma_0$ is $\sigma_0=\frac{3\lambda_0^2}{2\pi}=6.5\times10^{-14}~\mathrm{m^2}$, and natural line width $\Gamma$ is $\Gamma=20~ $MHz \cite{Olmschenk2009}.
In the following calculations, we assume that the cross-section of atom distribution in the probe beam $A_p$ is
$A_p=\pi (200~\mathrm{\mu m})^2=1.3\times 10^{-7}~\mathrm{m^2}$ and the atom number $N_a$ is $N_a=10^6$.  For these parameters, on-resonance optical depth OD is OD$\equiv N_a\sigma_0/A_p=0.52$.
We set the Zeeman splitting $\delta$ to be $\delta=\Gamma$.  This detuning corresponds to applying 14.3 Gauss to trapped ions.

We aim to measure the spin phase $\phi$, which is related to $\emph{w}$ at step (5) of the APL, as
\begin{equation}\label{}
    \phi=\arcsin\emph{w}=\arcsin(\frac{2N_\uparrow}{N_a}-1),
\end{equation}
where $N_{\uparrow}$ is the number of atoms with spin up and is defined as $N_\uparrow\equiv\frac{N_a}{2}(\emph{w}+1)$.
We need to be careful with the concept of $N_\uparrow$.  Each atom is in a superposition state, not in the eigenstate, so $N_{\uparrow}$ means ``the expected number of atoms with spin up if it were projected'' and has the uncertainty given by quantum projection noise.

\subsection{Faraday rotation with atoms}
We will estimate the Faraday rotation angle and generally follow the notation in Ref. \cite{Geremia2006} in this section.
The phase shift for each circularly polarized mode ($\sigma_\pm$) by a $|\!\uparrow\rangle$ atom can be written as
\begin{equation}\label{eq:varphi}
    \varphi_\pm=-\frac{\sigma_0}{A_p}\frac{\Delta_\pm/\Gamma}{1+(2\Delta_\pm/\Gamma)^2}.
\end{equation}
Here we modeled the resonance with the Lorentzian spectrum.
When $\varphi_+$ and $\varphi_-$ are different, the polarization plane of a linearly polarized wave is rotated.
This effect is known as Faraday rotation.
The evolution operator of the Faraday rotation is expressed as a combination of the phase shift operators (see \ref{App:Operators}),
\begin{eqnarray}
U'_\mathrm{FR}&=&\exp(i\varphi_+ n_+)\exp(i\varphi_-n_-)\nonumber \\
&=&\exp\left[i(\varphi_++\varphi_-)(n_++n_-)/2\right]U_\mathrm{FR},
\end{eqnarray}
where we define
\begin{equation}
U_\mathrm{FR} \equiv\exp\left(-i\gamma_z (n_+-n_-)/2\right). \label{U_FR}
\end{equation}
and $\gamma_z\equiv -(\varphi_+-\varphi_-)$.
Since the common shift between $\sigma_\pm$ modes does not affect the rotation angle,
 we neglect the factor between $U'_\mathrm{FR}$ and $U_\mathrm{FR}$
 and use Eq. (\ref{U_FR}) as the evolution operator of the Faraday rotation in the following.

By using Eq. (\ref{eq:varphi})
, $\gamma_z$ becomes
\begin{eqnarray}
\gamma_z &=& 2\beta N_\uparrow=\beta N_a(1+w),\\
\beta &\equiv& \left(\frac{\Delta_+/\Gamma}{1+(2\Delta_+/\Gamma)^2}-\frac{\Delta_-/\Gamma}{1+(2\Delta_-/\Gamma)^2}\right)\frac{\sigma_0}{2A_p}.
\end{eqnarray}
Note that $\beta N_a$ is a value of $\gamma_z$ when the atoms are equally populated to the $|\!\uparrow\rangle$ state and $|\!\downarrow\rangle$ state.
For our projected experimental condition (see Section \ref{Decoherence}
) of $\Delta=0$, i.e., $\Delta_+=-\Gamma$ and $\Delta_-=\Gamma$, $\beta$ is estimated to be  $\beta=-1.0\times10^{-7}$ rad/atom.

For convenience in the following discussions, we introduce a \textit{general} rotation operator of the polarization plane as
\begin{equation}
R(\theta)\equiv \exp\left(i\theta(n_+-n_-)\right), \label{R(theta)},
\end{equation}
 where $\theta$ is equal to the rotation angle of the linear polarization plane (see \ref{App:Operators}).
 $U_\mathrm{FR}=R(-\gamma_z/2)$, and the rotation angle of $U_\mathrm{FR}$ is $-\gamma_z/2$.

\subsection{Signal of polarimeter}
The rotation angle is measured using a polarimeter that outputs the photon number
difference between two orthogonal linearly polarized modes.
When we set these modes $h$ and $v$, the output of the polarimeter becomes
\begin{equation}
Q=\eta e(n_h-n_v),
\end{equation}
 where $\eta$ is the quantum efficiency of the detector.
Since the photons are randomly branched at the polarizing beam splitter,
we need a finite photon number to distinguish in which direction and to what angle the polarization is rotated.
To obtain the change from $w=0$,
the polarization angle of the incident pulse is set to $\pi/4+\beta N_a/2$
from the $h$-axis.
\begin{equation}
|\psi(0)\rangle=R(\pi/4+\beta N_a/2)|\alpha\rangle_h|0\rangle_v,
\end{equation}
where $|\alpha\rangle_h$ is a coherent state of $h$-mode
with a complex amplitude of $\alpha$. Note that $|\alpha|^2$ is the mean number of photons in a probe beam pulse.

\begin{figure}[htbp]
  \begin{center}
    \includegraphics[width=60mm]{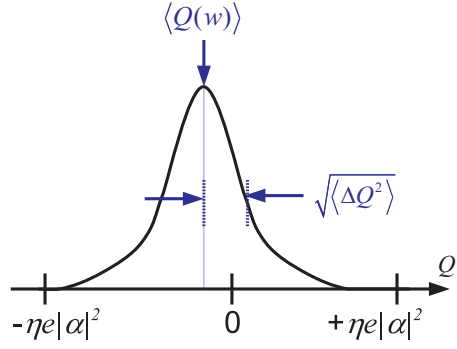}
  \end{center}
  \caption{Shot noise $\sqrt{\langle\Delta Q^2\rangle}$ and Faraday rotation signal $\langle Q(w)\rangle$. When $\langle Q(w)\rangle < \sqrt{\langle\Delta Q^2\rangle}$, the signal is not distinguished from 0.}
  \label{fig:shotnoise}
\end{figure}

The pulse after the transmission becomes
\begin{equation}
|\psi(w)\rangle
=U_\mathrm{FR}|\psi(0)\rangle
=R\left(\pi/4-\beta N_a w/2\right)|\alpha\rangle_h|0\rangle_v.
\end{equation}
The mean value, the square mean value, and the variance
 of the error signal become
\begin{eqnarray}
\langle Q(w)\rangle=&\eta e|\alpha|^2\sin(\beta N_a w),\\
\langle Q^2(w)\rangle=&\eta^2 e^2|\alpha|^2\left(1+|\alpha|^2\sin^2(\beta N_a w)\right),\\
\langle\Delta Q^2\rangle=&\langle Q^2(w)\rangle-\langle Q(w)\rangle^2=\eta^2 e^2|\alpha|^2. \label{shotnoise}
\end{eqnarray}
Since the average of the error signal is proportional to the $w$ component  for $\beta N_a w\ll 1$,
 $\langle Q(w)\rangle$ can be used as an error signal.
Eq. (\ref{shotnoise}) is known as the shot noise.
Note that we neglect the quantum projection noise in a collective spin system,
 in Eq. (\ref{shotnoise}).
If the spin noise were concerned,
 $w$ should be treated as a q-number and the variance would become
 $\langle\Delta Q^2\rangle=\eta^2 e^2|\alpha|^2(1+\beta^2 N_a^2 |\alpha|^2 \langle \Delta {w}^2\rangle)$.
If the photon number is large enough to satisfy $\langle Q(w)\rangle^2 > \langle\Delta Q^2\rangle$,
 we can recognize that the unbalance from $Q=0$ is caused by the atomic dephasing of $w$ rather than the shot noise.
For example,
\begin{equation}
|\alpha|^2 > \frac{1}{\sin^2(\beta N_a/10)}
\end{equation}
is required for the sensitivity of $w=0.1$ (see \Fref{fig:shotnoise}).
We define the photon number $N_{ph}(\rm{SNR})$ as a number of photons that satisfies $\langle Q(\frac{1}{\rm{SNR}})\rangle^2=\langle\Delta Q^2\rangle$, so
\begin{equation}\label{}
 N_{ph}(\rm{SNR})\equiv\frac{1}{\sin^2(\beta N_a /\rm{SNR})}.
\end{equation}
For our condition with $\Delta=0$, $N_a=10^6$, and SNR$=10$, $N_{ph}(10)=9.3\times10^3$ photons.

\subsection{\label{Decoherence}Decoherence}
There are three types of decoherence mechanism: spontaneous emission, light shift due to a probe beam, and backaction.  The dominant decoherence is due to spontaneous emission and is discussed here.  Decoherence due to light shift and backaction is discussed in \ref{app:backaction}.

For a linearly polarized pulse with a complex amplitude $\alpha$, $\langle n_\pm\rangle =|\alpha|^2/2$.
The total absorption probability by both circularly polarized modes for an atom due to a pulse illumination is
\begin{equation}
P_a=\sum_{i=\pm}\frac{\sigma_0/A_p}{1+(2\Delta_\pm/\Gamma)^2}\langle n_\pm\rangle=\varepsilon |\alpha|^2 \label{eq:P_a},
\end{equation}
where we set
\begin{equation}
\varepsilon=\left(\frac{1}{1+(2\Delta_+/\Gamma)^2}+\frac{1}{1+(2\Delta_-/\Gamma)^2}\right)\frac{\sigma_0}{2A_p}.
\end{equation}
For $\Delta=0$, $\varepsilon=1.0\times10^{-7}$.

Once an atom absorbs the photon, it will spontaneously emit the photon and is projected to $|\!\uparrow\rangle$.  This is the main source of decoherence with our experimental setup.  In APL, we repeat the measurement with $N_{ph}(\rm{SNR})$ for $N_{rep}$ times.  The total probability of decoherence is given from the total number of photons, as in Eq. (\ref{eq:P_a}), and we obtain
\begin{equation}\label{}
    P_a^{total}(\rm{SNR})=\varepsilon N_{ph}(\rm{SNR}) N_{rep}.
\end{equation}
Decoherence is safely permitted to be $\frac{1}{\rm{SNR}}$, so $N_{rep}$ is given by
\begin{equation}\label{eq:Nrep}
    N_{rep}=\frac{1}{\varepsilon\ N_{ph}(\rm{SNR})\ \rm{SNR}}.
\end{equation}

\begin{figure}[htbp]
  \begin{center}
    \includegraphics[width=80mm]{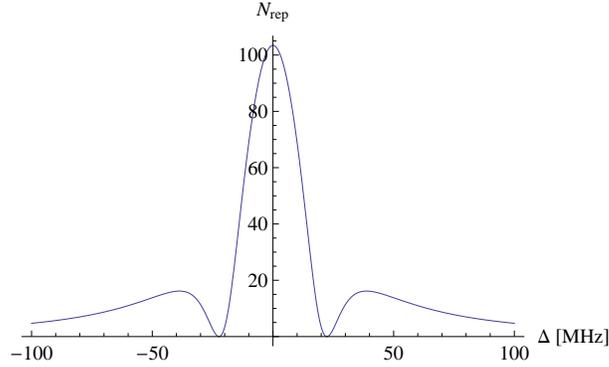}
  \end{center}
  \caption{$N_{rep}$ as a function of $\Delta$.  $N_a=10^6$, SNR=10, $\Gamma=20$ MHz, $A_p=1.3\times10^{-7}$ m$^2$, and $\sigma_0=6.5\times10^{-14}$.}
  \label{fig:Nrep}
\end{figure}

$N_{rep}$ as a function of $\Delta$ is shown in \Fref{fig:Nrep}.
We can see that $N_{rep}$ is maximum at $\Delta=0$ and has the value of $1.1\times10^2$ when SNR$=10$.  This means that we can repeat APL for 100 cycles.

Note that at $\Delta=0$, the probe beam is resonant to a $^2S_{1/2}(F=1, m_F=0)$ $\rightarrow$ $^2P_{1/2}(F=1, m_F=0)$ transition, but this transition is forbidden because $\Delta F=\Delta m_F=0$.

\section{\label{NumericalSimulation}Numerical simulation}
To confirm that the Allan deviation is reduced at rate $\tau^{-1}$, we ran a simple numerical simulation.  We first generated a free-run LO noise of the flicker frequency noise type.  We used a noise level of $\sigma_y^{free run}(\tau)=1\times10^{-12}$ to model the noise of a crystal oscillator.

The feedback schematics are shown in \Fref{fig:ExperimentalSetup} (b).

For the projection Ramsey method, we applied the following feedback scheme:
\begin{equation}\label{eq:FeedbackProjection}
    y^{(i)}_{projection}=y^{(i-1)}_{projection}-G_P\frac{\Delta\phi^{(i)}}{2\pi \nu_0{T_{FP}}},
\end{equation}
where $y^{(i)}_{projection}$ is the updated normalized frequency at the end of the $i$-th cycle, $G_P$ is the ``proportional'' gain, and $\Delta\phi^{(i)}$ is the phase accumulated during the $i$-th cycle.  We modeled the effect of projection measurement by resetting the phase at every cycle.

For APL, feedback frequency is given by
\begin{equation}\label{eq:FeedbackSCM}
    y^{(i)}_{APL}=y^{(i-1)}_{APL}-G_P\frac{\phi^{(i)}-\phi^{(i-1)}}{2 \pi \nu_0 T_{FP}},
\end{equation}
where $\phi^{(i)}$ is the (total) phase measured via weak measurement. Note that this $\phi^{(i)}$ cannot be obtained unless the phase is maintained over $(i)$ cycles.
The measurement noise that corresponds to $\sigma^Q$ of the given SNR is included in $\Delta\phi$ or $\phi$.
We assumed that the dead time was zero for both projection Ramsey and APL.
For APL, we included the effect of the decoherence by resetting the phase every 100 cycles.

\begin{figure}[htbp]
  \begin{center}
    \includegraphics[width=100mm]{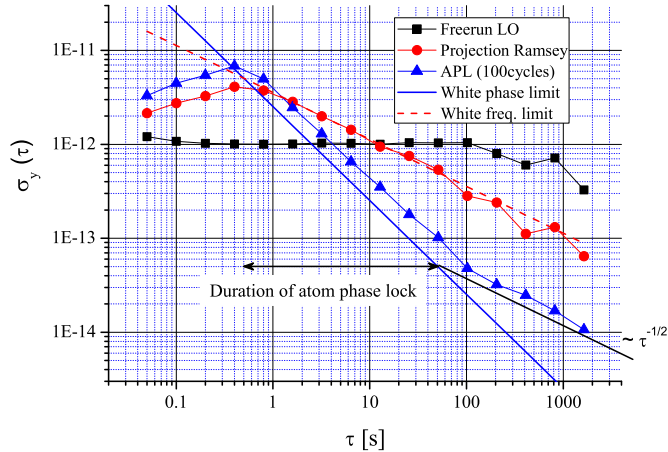}
  \end{center}
  \caption{Numerical simulation of the Allan deviation for the projection Ramsey method, APL, and free-run LO.  The phase of the spin is reset every 100 cycles (50 s). Step time width $dt=50$ ms, carrier frequency $\nu_0=12.6$ GHz, SNR=10, cycle time $T_c=T_{FP}=500$ ms, and proportional gain $G_P=1$.}
  \label{fig:Simulation100cycles}
\end{figure}

The results of the numerical simulation are shown in \Fref{fig:Simulation100cycles}.
This graph show that the Allan deviation is averaged down at a rate of $\propto \tau^{-1/2}$ for a feedback cycle with the projection Ramsey method.
For APL, the stability decreases as $\tau^{-1}$, from $\tau=0.5$ to 50 s.  The phase is reset every 50 s (100 measurement cycles), and thus the stability decreases as $\tau^{-1/2}$  from 50 s, which is the same slope as that of the projection Ramsey method.

SNR=10 was chosen to minimize the decoherence rate during APL.  For comparison, the SNR of the projection Ramsey method was also set to 10, representing a case when SNR is limited by technical noise \cite{Itano1993} at the 10 $\%$ level.

\section{Discussion}
For APL to achieve white phase noise, we need to trap a large number of ions in an optical path and also keep the dead time sufficiently small compared to $T_{FP}$.  Dead time is the time taken by steps (2), (4), (5), and (6) in \Fref{fig:AtomPhaseLock}.
For the number of atoms, we chose $N_a=10^6$ to have a decent OD.  If we try to confine ions in $A_p=1.3\times 10^{-7}\ \rm{m}^2$ and assume the inter-distance of ions to be 30 $\mu$m, then the trap length needs to be at least 20 cm to have $10^6$ ions.  We can avoid making a long trap by forming a bad cavity along the probe beam, around the ions of finesse $\sim$100, and reduce the number of ions to $10^4$ and the size of the trap to 2 mm.
As for the dead time, its effect in a case with projection noise has been thoroughly investigated by Dick \cite{Dick1987}.  For APL, we need to at least keep the dead time low enough not to be limited by the so-called Dick limit. In other words, APL will work only as long as the noise accumulated during the dead time is negligible.  A quantitative discussion would require careful evaluation and simulation,  but this is beyond the scope of this paper.

We have discussed the stability of APL based on a simulation  assuming that atomic clock frequency does not change over time.  When atomic clock frequency is unstable, the stability of the atomic clock will be limited by the stability of the atomic clock frequency itself.
The systematic uncertainty of the $^{171}\rm{Yb}^+$ ions, cooled by helium buffer gas, is reported to be 6 mHz \cite{Tamm1995}.  Even with this rather high systematic uncertainty, we can still expect the Allan deviation to decrease to $\sigma_y=5\times10^{-13}$ or lower without further cooling.

The experimental setup for APL coincides with that of spin squeezing.   Wineland proposed using the spin-squeezed state to overcome quantum projection noise \cite{Wineland1992, Wineland1994}, and spin squeezing for microwave clock transitions has been demonstrated \cite{Louchet-Chauvet2010, Leroux2010}.  The comparison between APL and spin squeezing is an interesting topic and will be published elsewhere.  Here, we briefly mention that both use the weak (dispersion) measurement, but squeezing aims to improve the SNR in one measurement while APL initially uses a lower SNR but aims to improve the stability over many cycles by preserving the phase longer and avoid the QPN limit in the long term.

In a similar spirit, to improve the (short term) stability of the atomic clock, an optical active clock has been proposed \cite{Chen2009, Meiser2009}.  However, the design of the optical active clock consists of an optical cavity, and long-term stability is again limited by the thermal noise of the mirror coating.


Note that APL primarily contributes to improving the stability, and not directly to the accuracy of the atomic clock.  However, improvement of the stability would enable the experimental estimation of the accuracy to be much shorter and indirectly contribute to improving the accuracy.

Once APL is demonstrated with the microwave clock, the same principle should be applied to the neutral atom clock trapped in the optical lattice.  We propose to use rf-ion traps for proof-of-principle demonstration with a microwave clock, but there will be difficulties in applying APL to such traps.  First, buffer gas cooling will no longer be sufficient for ions to stay within the Lamb-Dicke limit because the wavelength is much shorter.  Second, AC stark shift due to a trap rf-field will shift the optical clock frequency.  An optical lattice clock is a candidate for avoiding those two problems.  The time scale for applying APL will be shorter (up to $\sim$1 sec) because the trapping time of the optical lattice clock is short ($\sim$1sec) compared to an rf-ion trap ($\sim$days).

\ack
This work was supported by the JST PRESTO program.  We thank N. Yamamoto and K. Sugiyama for fruitful discussions, and S. Nagano and A. Yamaguchi for comments on this paper.

\appendix
\section{Phase shift operator and rotation operator of polarization plane}\label{App:Operators}
An evolution operator that gives the phase shift is $\exp(i\varphi_\pm n_\pm)$, where $n_\pm$ is the number operator for $\sigma_\pm$ mode.
This can be seen by the following equation.
After unitary conversions with the phase shift operators, the photon-annihilation operator for $\sigma_\pm$ mode, say $a_\pm$, becomes
\begin{equation}
\exp(-i\varphi_\pm n_\pm)a_\pm\exp(i\varphi_\pm n_\pm)=a_\pm e^{i\varphi_\pm}, \label{PSOperator}
\end{equation}
where the commutation relation between the number and annihilation operator $[n_\pm,a_\pm]=-a_\pm$,
and the Baker-Hausdorff lemma $\exp(\lambda B)A\exp(-\lambda B)=A+\lambda[B,A]+(\lambda^2/2!)\left[B,\right[B,A]]+\cdots$
are used.
Since the annihilation operator in quantum optics corresponds to the amplitude of the  EM field in classical electrodynamics,
it can be said that the phase shift operator $\exp(i\varphi_\pm n_\pm)$ gives the phase shift.

An evolution operator that gives the rotation of polarization plane takes the form of Eq. (\ref{R(theta)}).
This can be seen by the following equation.
We introduce two linear polarization modes $h$ and $v$, which are related to the $\sigma_\pm$ mode, as
\begin{equation}
a_\pm=\frac{a_h\pm ia_v}{\sqrt{2}}.
\end{equation}
After the unitary conversion with the rotation operator $R(\theta)$, the photon-annihilation operator $a_{h,v}$ becomes
\begin{eqnarray}
R^\dagger(\theta)a_hR(\theta)&=&\frac{a_+e^{i\theta}+a_-e^{-i\theta}}{\sqrt{2}}=a_h\cos\theta-a_v\sin\theta,\\
R^\dagger(\theta)a_vR(\theta)&=&\frac{a_+e^{i\theta}-a_-e^{-i\theta}}{\sqrt{2}i}=a_h\sin\theta+a_v\cos\theta,
\end{eqnarray}
where Eq. (\ref{PSOperator}) is used.
Thus, it can be said that the rotation angle of the polarization plane is $\theta$.

\section{Light shift and backaction}\label{app:backaction}
The probe beam affects the spin state in two ways: light shift (AC Zeeman shift) and imbalance fluctuation of the right and left circularly polarized component.  We call the latter case ``backaction.''
Both cases rotate the spin along the longitudinal $u-v$ direction, but the difference is that light shift rotates the spin direction at a constant rate while backaction rotates randomly, resulting in increase of the deviation along the longitudinal direction.
Light shift is zero for the case with $\Delta=0$ because the contributions from $^2P_{1/2}(F=0, m_F=\pm1)$ cancel each other.
In the following, we will show that the effect of backaction is also negligible compared to the decoherence due to absorption, which has already been discussed.

To estimate the backaction, we need to treat the number of up spins (represented by $w$) as a q-number.  For this purpose, we need to treat $(u,v,w)$ as collective spin operators \cite{Kitagawa1993}.
\begin{eqnarray}
\hat{u} &=&\frac{1}{N_a}\sum_{i=1}^{N_a}\left(|g,\uparrow\rangle_i{}_i\langle g,\downarrow|+|g,\downarrow\rangle_i{}_i\langle g,\uparrow|\right)\\
\hat{v} &=&\frac{i}{N_a}\sum_{i=1}^{N_a}\left(-|g,\uparrow\rangle_i{}_i\langle g,\downarrow|+|g,\downarrow\rangle_i{}_i\langle g,\uparrow|\right)\\
\hat{w} &=&\frac{1}{N_a}\sum_{i=1}^{N_a}\left(|g,\uparrow\rangle_i{}_i\langle g,\uparrow|-|g,\downarrow\rangle_i{}_i\langle g,\downarrow|\right),
\end{eqnarray}
where $|g,\uparrow\rangle_i$ ($|g,\downarrow\rangle_i$) represents the state that the $i$-th atom is $|g,\uparrow\rangle$ ($|g,\downarrow\rangle$).
Their commutation relations are as follows:
\begin{equation}
[\hat{u},\hat{v}]=\frac{2i}{N_a}\hat{w},\quad [\hat{v},\hat{w}]=\frac{2i}{N_a}\hat{u},\quad [\hat{w},\hat{u}]=\frac{2i}{N_a}\hat{v}.
\end{equation}
From one of the commutation relations, we obtain an uncertainty relation
\begin{equation}\label{eq:QPN}
\langle\Delta\hat{v}^2\rangle\langle\Delta\hat{w}^2\rangle\ge\frac{|\langle\hat{u}\rangle|^2}{N_a^2}.
\end{equation}
In the following discussion, we consider $\langle\hat{u}\rangle=\langle\hat{u}^2\rangle=1$.
The minimum uncertainty state $\langle\Delta\hat{v}^2\rangle=\langle\Delta\hat{w}^2\rangle=1/N_a$
 corresponds to the quantum projection noise limit.

 After a probe pulse illumination,
the atomic state operator evolves from $\hat{A}$ to $\hat{A}'$ as
\begin{equation}
\hat{A}'=\langle\psi(0)|\hat{U}_\mathrm{FR}^\dagger\hat{A}\hat{U}_\mathrm{FR}|\psi(0)\rangle,
\end{equation}
where
\begin{equation}
\hat{U}_\mathrm{FR}=\exp\left(-i\beta N_a(1+\hat{w})(n_+-n_-)\right)
\end{equation}
is the evolution operator of the Faraday rotation treating $w$ as a q-number.
By using the equations of motion under a fictitious magnetic field,
\begin{eqnarray}
\hat{U}_\mathrm{FR}^\dagger\hat{u} \hat{U}_\mathrm{FR}
&=&\hat{u}\cos\left(2\beta(n_+-n_-)\right)-\hat{v}\sin\left(2\beta(n_+-n_-)\right),\\
\hat{U}_\mathrm{FR}^\dagger\hat{v} \hat{U}_\mathrm{FR}
&=&\hat{u}\sin\left(2\beta(n_+-n_-)\right)+\hat{v}\cos\left(2\beta(n_+-n_-)\right),\\
\hat{U}_\mathrm{FR}^\dagger\hat{w} \hat{U}_\mathrm{FR}
&=&\hat{w},
\end{eqnarray}
and the fluctuation of the circularly polarized component for linearly polarized coherent light
\begin{eqnarray}
\langle\psi(0)|(n_+-n_-)|\psi(0)\rangle&=&0,\\
\langle\psi(0)|(n_+-n_-)^2|\psi(0)\rangle&=&|\alpha|^2,
\end{eqnarray}
we obtain the increase of the variance
\begin{eqnarray}
\langle\Delta\hat{u}'^2\rangle
&=&4\beta^2|\alpha|^2\langle\hat{v}^2\rangle\\
\langle\Delta\hat{v}'^2\rangle
&=&\langle\Delta\hat{v}^2\rangle\left(1-4\beta^2|\alpha|^2\right)+4\beta^2|\alpha|^2 \label{backaction}.
\end{eqnarray}
In the second term of Eq. (\ref{backaction}) on the right side, $4\beta^2|\alpha|^2$ represents the backaction noise.  For our experimental parameters, variance is increased along the longitudinal direction by $4\beta^2|\alpha|^2=3.7\times10^{-10}$ per measurement of SNR=10.
The backaction is negligible under the condition of $4\beta^2|\alpha|^2N_a\ll 1$.
This condition is satisfied as long as the sensitivity of the Faraday rotation is well below the quantum projection noise limit.

\section*{References}
\bibliographystyle{unsrt}

\end{document}